\documentclass[twocolumn,preprintnumbers]{revtex4-1}
\usepackage{amssymb}
\usepackage{mathrsfs}
\usepackage{float}
\usepackage{bm}
\usepackage{amsmath}
\usepackage{mathdots}
\usepackage{graphicx}
\usepackage{subfigure}
\usepackage{amsmath}
\usepackage{array}
\usepackage{lipsum}

\begin{document}
\newcommand*{\PKU}{School of Physics, Peking University, Beijing 100871,
China}\affiliation{\PKU}

\title{Testing the Higgs sector directly in the nonrelativistic domain}

\author{Zhentao Zhang}\email{zhangzt@pku.edu.cn}\affiliation{\PKU}

\begin{abstract}
Directly measuring the Higgs self-coupling is of great importance for testing the Brout-Englert-Higgs mechanism in the Standard Model. As a scattering that contains the direct information from the Higgs self-coupling, we investigate the process $\mu^-\mu^+\rightarrow HH$ in the threshold region. We calculate the one-loop corrections to the cross section and consider the non-perturbative contribution from the Higgs self-interactions in the final state. It is found that the scattering in the nonrelativistic domain could be an especial process to testing the Higgs sector directly.
\end{abstract}
 \maketitle

\section{Introduction}
In the Standard Model (SM), it is nearly impossible to overstate the importance of the Brout-Englert-Higgs (BEH) mechanism. Recently, the Higgs boson predicted by the BEH mechanism was observed at the CERN Large Hadron Collider~\cite{ATLAS,CMS}. It is undoubted that the discovery of the last particle in the SM is a great success for the model. However, the existences of neutrino oscillations and dark matter remind us that there is new physics beyond the model. To find hints for the new physics beyond the SM, we need to strictly test the Higgs sector in the SM, and the only way to reconstruct the details of the Higgs sector is to precisely determine the Higgs self-couplings.

A few decades ago, it was found that the muon-induced Higgs physics could offer a significant way for precisely probing the Higgs sector~\cite{Barger,Atwood,Pilaftsis,Choi}. In this paper, we shall study the process $\mu^-\mu^+\rightarrow HH$ in the threshold region, since this muon-induced process contains the direct information from the Higgs self-coupling. We know that in perturbation theory, the calculation for the one-loop cross section is well-established. However, near the threshold the situation would become subtle when we want to obtain the accurate information for the cross section. After dealing with this issue, we shall show that this process in the threshold region has the advantages to be a valuable way to testing the Higgs sector directly.

\section{Threshold Higgs pair production}
It has been known for a long time that appearance of a static potential between two nonrelativistic particles in the initial (final) state may significantly distort the incoming (outgoing) wave-function for the plane wave approximation in scattering theory~\cite{Sommerfeld}, and then due to the nonrelativistic interactions, we need to consider the corrections for the scattering wave-function in the reaction zone. This Sommerfeld effect was widely considered in many branches of quantum physics. In recent years, the extensive studies of dark matter annihilation have revived this effect for short-range forces~\cite{Hisano,Cirelli,Arkani-Hamed,Lattanzi,March-Russell,Cassel,Slatyer,Zhang1}, and the non-perturbative nature of this effect was stressed. We may note that in the nonrelativistic domain, even though there is no zero-energy resonance or strong enhancement which is obviously beyond the content of the purely perturbative calculation, in general the convergence of the ladder diagrams for the process would be a lot slower than it at high energies. Thus, to obtain more reliable results for the threshold Higgs pair production, we need to consider this non-perturbative effect to the cross section.

There are three diagrams that contribute to the process at the leading order. However, an elementary calculation can show that the contribution of the $u$- and $t$-channels compared to the $s$-channel are extremely small, and we only need to consider the $s$-channel contribution in the Born amplitude. At the next-to-leading order, many more diagrams would contribute to the cross section in the SM. Since the $u$- and $t$-channels contribution can be safely ignored in the Born amplitude and the scattering wave-function would be distorted intensely only near the origin, it can be understood easily that in the loop diagrams we could also ignore the contribution from the final state interactions in the $u$- and $t$-channels.

The system of the Higgs boson pair produced in the threshold region is nonrelativistic and any nonrelativistic
interactions arisen in the system could modify the probability amplitude for the final state.
The cross section including the final state interactions is
\begin{equation}\label{simpleforFSI}
\sigma_\text{H}=|\psi(0)|^2\sigma_\text{Born},
\end{equation}
where the correction term $|\psi(0)|^2$ is the well-known Sommerfeld factor, and $\psi(r)$ is the Schr\"{o}dinger wave function for the two-body system.

We know that in nonrelativistic quantum mechanics the Schr\"{o}dinger wave function in a central field is
\begin{equation}
\psi=\frac{1}{2k}\sum\limits_{l = 0}^\infty  {{i^l}}( {2l + 1}){e^{i{\delta _l}}}{P_l}( {\cos \theta}){R_{kl}}(r),
\end{equation}
where ${\delta _l}$ are the phase shifts and ${R_{kl}}(r)$ are the radial functions. The $R_{kl}$ are normalized as $\int^{\infty}_{0}r^2R_{k'l}R_{kl}dr=2\pi\delta(k'-k)$. Thus
\begin{equation}\label{Troublemaker}
|\psi(0)|^2=|R_0(0)/2k|^2=|\phi(r)/2kr|^2_{r\rightarrow0},
\end{equation}
where $R_0$ is the s-wave radial function, and $\phi(r)=rR_0(r)$.

The wave-function $\phi(r)$ satisfies the reduced s-wave two-body radial Schr\"{o}dinger equation
\begin{equation}\label{s-wave}
  \frac{{d^2}\phi(r)}{d{r^2}} - m_{H}V({r})\phi(r) =-{(m_{H}\beta)^2}\phi(r),
\end{equation}
where $\beta$ is the velocity of the Higgs bosons and $m_{H}$ is the mass of the Higgs boson.

To find the wave-function $\phi(r)$, we next need to know the forces between the two nonrelativistic Higgs bosons. We know that in the SM the Higgs sector after the electroweak symmetry breaking can be written in the form $V(H)=\frac{1}{2}{m^2_H}{H}^2+\lambda{v}{H}^3+\frac{1}{4}\lambda H^4$, where $v$ is the vacuum expectation value and $\lambda$ is the Higgs self-coupling constant. Without considering any possible interactions beyond SM, the nonrelativistic potential for the Higgs self-interactions may be established from the process $HH\rightarrow HH$ as~\cite{Zhang2}
\begin{align}\label{all}
  V(\bm{r}) =\frac{3\lambda}{m^2_H}\delta^{(3)}(\bm{r})-\frac{\alpha}{ r}e^{-m_Hr},
\end{align}
where coupling constant $\alpha=9\lambda/(8\pi)$. Then we are able to solve the radial Schr\"{o}dinger equation numerically with proper boundary conditions.

Since the Higgs correction factor $|\psi(0)|^2$ would be calculated non-perturbatively and the other loops are ``hard'' (i.e., short-distance) radiative corrections, $|\psi(0)|^2$ could also be counted as the correction to the final-state wave functions for their amplitudes. Therefore, after considering the ``hard'' radiative corrections at one-loop, we may express the cross section including the non-perturbative corrections as
\begin{equation}\label{full}
\sigma=|\psi(0)|^2(\sigma_{\text{Born}}+\sigma_{\text{R}}),
\end{equation}
where $\sigma_{\text{R}}$ is the ``hard'' radiative correction. It may be expected that the cross section $\sigma$ in Eq.~(\ref{full}) would be larger than the one-loop perturbative result.

Note that if we want to find the non-perturbative corrections from the Higgs potential in Eq.~(\ref{all}), we just need to consider the Yukawa part, and the $\delta$ potential
\begin{equation}\label{delta}
  V_\delta=\frac{3\lambda}{m^2_H}\delta^{(3)}(\bm{r})
\end{equation}
need not to be included \cite{Zhang2}. However, we should note that in fact in the one-loop calculation it would not ignore the contribution of the $\delta$ potential.

\begin{figure}[H]
\centering
\includegraphics[width=4.5cm]{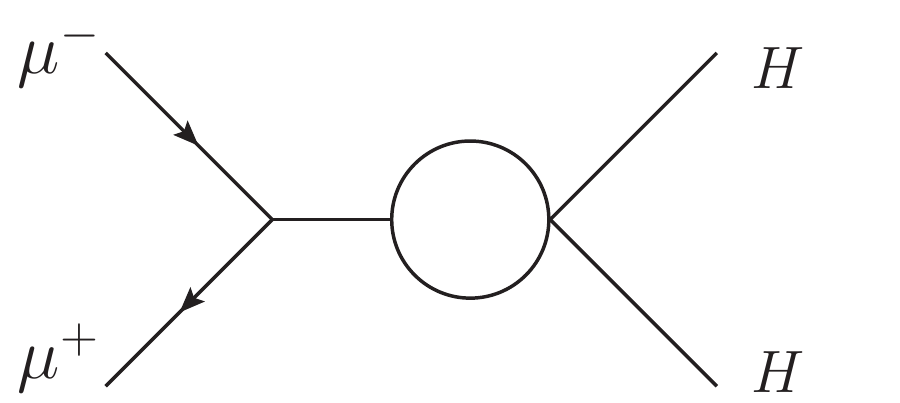}~
\includegraphics[width=4.5cm]{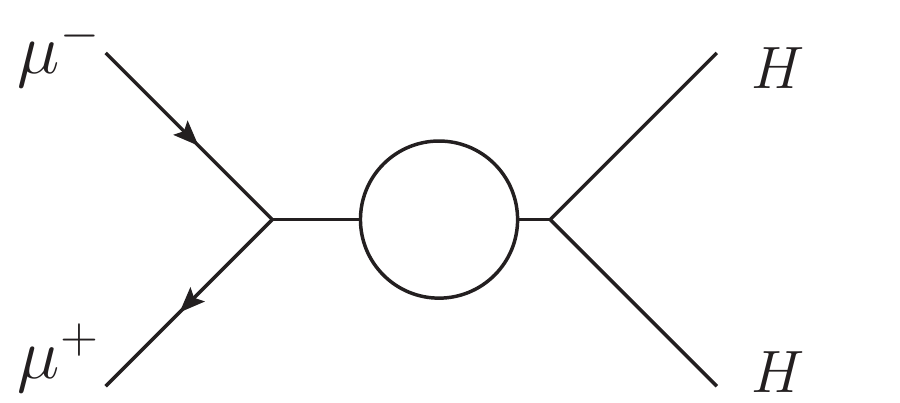}
 \caption{The diagrams correspond to the $\delta$ potential scattering in the $s$-channel. Solid line denotes Higgs boson.}\label{Loop}
\end{figure}

Notice that the $\delta$ potential is from the trilinear and quadrilinear Higgs self-couplings~\cite{Zhang2}. At the next-to-leading order we may find that the two relevant diagrams for the scattering are the diagrams shown in Fig.~\ref{Loop}. It can be seen that the $\delta$ potential comes from the diagrams when we put the Higgs bosons in the loop on mass-shell. Nevertheless, these diagrams may be treated perturbatively if we do not consider the higher order corrections. Therefore the contribution of the $\delta$ potential in Eq.~(\ref{delta}) would be counted automatically in the one-loop perturbative calculation.

\begin{figure}[H]
\centering
\includegraphics[width=9cm]{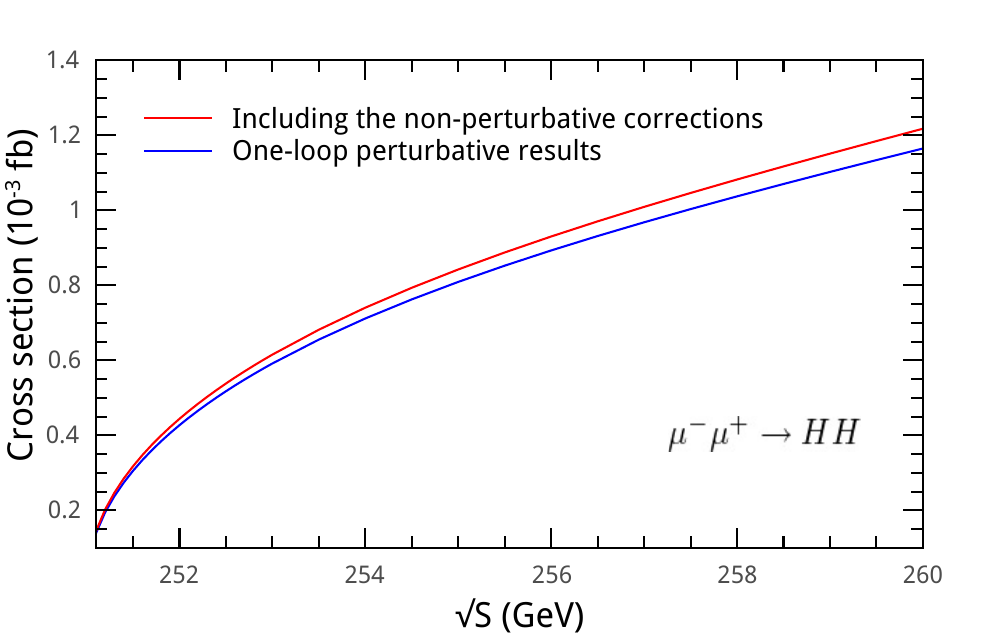}
   \caption{The cross sections in the threshold region. The blue one is one-loop cross section in perturbation theory, and the red one is calculated including the non-perturbative corrections. The energy range starts from $251.1$ GeV. }\label{Results}
\end{figure}

To find the one-loop cross section including the non-perturbative corrections from the SM Higgs sector, let us carry out the calculation. We note that the ``hard'' radiative corrections would be computed through the computer programs \textit{FeynArts}~\cite{Hahna} and \textit{FormCalc}~\cite{Hahnb}, and the mass of the Higgs boson is set as $125.5$ GeV. The results of the total cross section in the threshold region are presented in Fig.~\ref{Results}. Meanwhile, as a useful reference, we also present the one-loop perturbative results for this process.

\begin{figure}[H]
\centering
 \includegraphics[width=9cm]{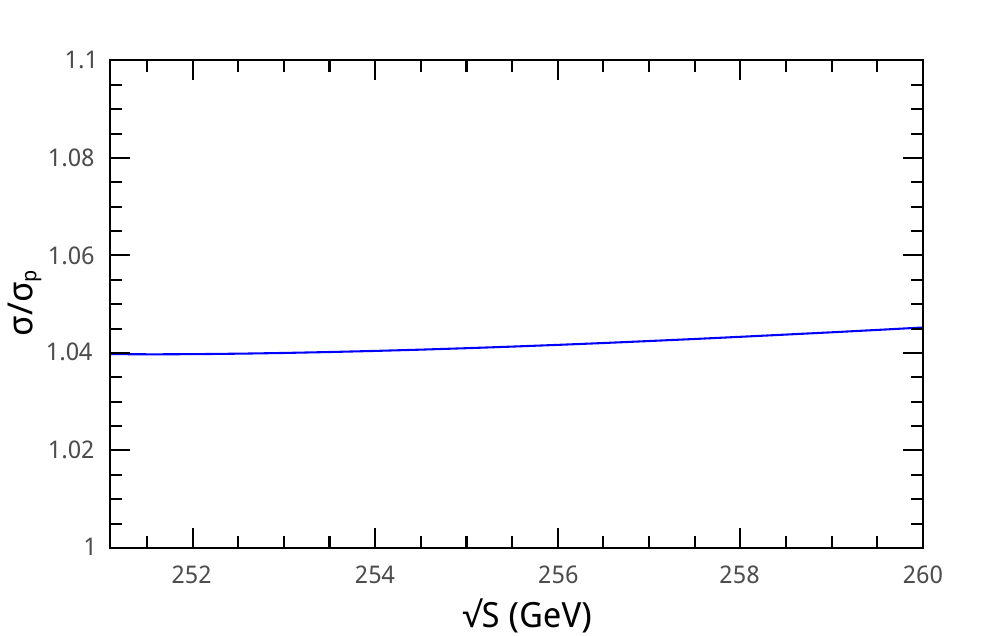}
 \caption{The Ratio of $\sigma/\sigma_{\text{p}}$ for the process $\mu^-\mu^+\rightarrow HH$ in the threshold region, where $\sigma_{\text{p}}$ is the one-loop cross section in perturbation theory.}\label{Ratio}
\end{figure}

It is found that the cross section including the non-perturbative corrections is always larger than the perturbative result, and in fact this is what we expect from the start. The reason is that in the nonrelativistic domain, through solving the Schr\"{o}dinger equation non-perturbatively, we obtain the contribution of the Higgs potential at all orders. For precisely measuring the Higgs self-coupling in the SM, it would be necessary for us to take into account the non-perturbative corrections, see Fig.~\ref{Ratio}.

Note that we may also find that close to the ``zero-momentum'' Higgs pair production region, the cross section decreases fast. This behavior is due to the fact that the cross section would be proportional to the velocity of the Higgs boson in the low-energy limit \cite{Zhang2}.

As is known to all, testing the Higgs sector through the direct measurement of the Higgs self-coupling is of great importance. As a process related directly to the Higgs self-coupling, two characteristics of the scattering $\mu^-\mu^+\rightarrow HH$ may deserve to be highlighted. First, we could easily obtain the accurate information of the Higgs self-coupling in the threshold scattering, since in the threshold region the main contribution to the cross section is from the s-channel Born amplitude and an overall factor comes from the Higgs self-interactions in the final state. Second, in the nonrelativistic domain it may be easier for us to find the evidence for the new physics hidden in the Higgs sector. Notice the fact that in the SM the non-perturbative effect in the final state arises from the Higgs self-interactions alone. If there are new particles in the Higgs sector, these particles may contribute to the effect too. Then through the non-perturbative effect the new physics in the threshold region could manifest itself more easily than it would be in perturbation theory. To get a better view on this point, let us elaborate on it in the following. The non-perturbative corrections from the Higgs boson itself are not dramatic, and however there is an interesting situation that a new particle which couples to the SM Higgs boson is light and the naively Feynman diagram expansions break down in the nonrelativistic domain, and then the non-perturbative corrections from the new particle would cause dramatic enhancement to the cross section. Specially, if we are lucky enough, the new particle lives in the parameter spaces of the zero-energy resonance in the scattering, and then it would make the observation of the new interactions is much easier because the zero-energy resonance could increase the cross section by several orders of magnitude. Note that with the help of the nonrelativistic scattering theory, we could give the general discussion without restricting ourselves to any specific new physics model.

\begin{figure}[H]
\centering
 \includegraphics[width=9cm]{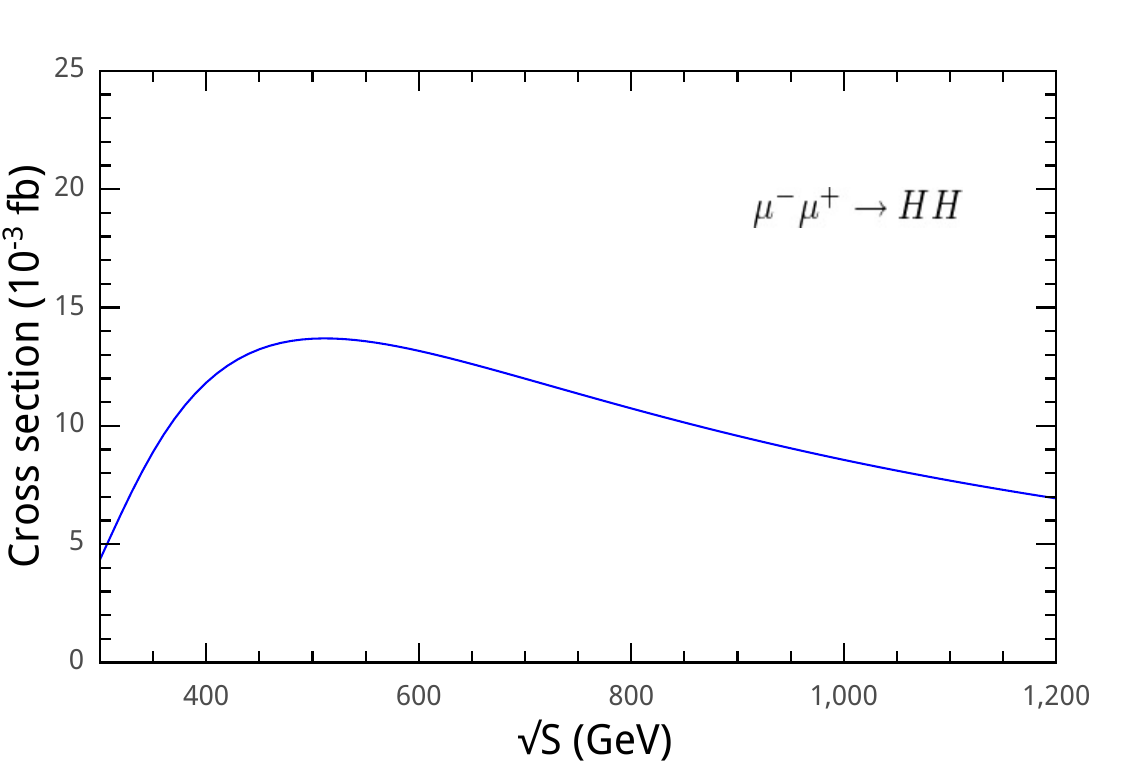}
\caption{The cross section away from the threshold region.}\label{Highenergy}
\end{figure}

Notice that in recent years progresses of the concepts and technologies for building high luminosity muon collider are constantly being made~\cite{Palmer,Bonesini}. It is evident that in the future it requires high luminosity muon colliders to detect the process in the threshold region, since the cross section for the low-energy Higgs pair production in the SM is small. However, we should note that even at high energies the cross section would only be about one order of magnitude larger than the threshold, see Fig.~\ref{Highenergy}. This fact implies that the detection of the process at high energies may be not less of a challenge than it in the threshold region. And the most important thing that should be underlined is that compared to the threshold scattering, extraction of the information from the Higgs self-coupling in the scattering would be very difficult, if not impossible, at high energies, because in this case the absolutely dominant contribution to the process would not come from the diagrams that contain the information of the Higgs self-coupling. Accordingly, although we do not consider the attractive fact that the non-perturbative corrections of any new interactions hidden in the Higgs sector would only appear at the threshold, to directly measure the Higgs self-coupling through the process $\mu^-\mu^+\rightarrow HH$, the threshold region is preferred.

\section{Summary}
After discovery of the Higgs boson, precisely testing the Higgs sector in the SM has great significance for particle physics. Since the process $\mu^-\mu^+\rightarrow HH$ contains the direct information from the Higgs self-coupling, in this paper, we present a detail analysis of the one-loop cross section for the scattering. Specially, we calculate the one-loop cross section including the non-perturbative contribution from the Higgs self-interactions in the final state, and this corrections may need to be considered for precisely measuring the Higgs self-coupling. We realize that it will be a challenge for us to detect the process. However, due to its attractive features, in the future this process in the threshold region may be considered as an especial way to testing the SM Higgs sector directly.

 \end{document}